\documentclass[%
reprint,
superscriptaddress,
amsmath,amssymb,
prl,
]{revtex4-2}
\usepackage{color}
\usepackage{float}
\usepackage[pdftex]{graphicx}
\usepackage{graphicx}% Include figure files
\usepackage{dcolumn}% Align table columns on decimal point
\usepackage{bm}% bold math
\usepackage{siunitx}
\usepackage{comment}
\usepackage{fixltx2e}
\usepackage{soul}
\usepackage{braket}

\usepackage{hyperref}% add hypertext capabilities
\hypersetup{
    bookmarks=true,         % show bookmarks bar?
    unicode=false,          % non-Latin characters in Acrobat’s bookmarks
    pdftoolbar=true,        % show Acrobat’s toolbar?
    pdfmenubar=true,        % show Acrobat’s menu?
    pdffitwindow=false,     % window fit to page when opened
    pdftitle={A single-electron spin qubit in a CMOS device},    % title
    pdfauthor={Dr. Matias Urdampilleta},     % author
    pdfsubject={},   % subject of the document
    pdfcreator={Dr. Matias Urdampilleta},   % creator of the document
    pdfproducer={},  % producer of the document
    pdfkeywords={,} {} {}, % list of keywords
    pdfnewwindow=true,      % links in new window
    colorlinks=false,       % false: boxed links; true: colored links
    linkcolor=red,          % color of internal links
    citecolor=green,        % color of links to bibliography
    filecolor=magenta,      % color of file links
    urlcolor=cyan           % color of external links
}
\usepackage{filecontents}
\usepackage{comment}

\DeclareUnicodeCharacter{2009}{\,} 
\DeclareSIUnit{\belmilliwatt}{Bm}
\DeclareSIUnit{\dBm}{\deci\belmilliwatt}

\begin{document}
\bibliographystyle{apsrev4-1}

\preprint{APS}
\title{Electrical manipulation of a single electron spin in CMOS using a micromagnet and spin-valley coupling}% Force line breaks with \\

% Alternative Title Bernhard
%\title{Valley-enhanced electrical manipulation of a single electron spin using post-CMOS processed micromagnet}% Force line breaks with \\

%\title{Electrical manipulation of a single electron spin in CMOS with micromagnet and spin-valley coupling}% Force line breaks with \\

\author{Bernhard Klemt$^\ast$}
\email{bernhard.klemt@neel.cnrs.fr}
\affiliation{Univ. Grenoble Alpes, CNRS, Grenoble INP, Institut N\'eel, 38402 Grenoble, France}

\author{Victor Elhomsy$^\ast$}
\affiliation{Univ. Grenoble Alpes, CNRS, Grenoble INP, Institut N\'eel, 38402 Grenoble, France}

\author{Martin Nurizzo}
\affiliation{Univ. Grenoble Alpes, CNRS, Grenoble INP, Institut N\'eel, 38402 Grenoble, France}

\author{Pierre Hamonic}
\affiliation{Univ. Grenoble Alpes, CNRS, Grenoble INP, Institut N\'eel, 38402 Grenoble, France}

\author{Biel Martinez}
\affiliation{Univ. Grenoble Alpes, CEA, IRIG, 38000 Grenoble, France}

\author{Bruna Cardoso Paz}
\affiliation{Univ. Grenoble Alpes, CNRS, Grenoble INP, Institut N\'eel, 38402 Grenoble, France}

\author{Cameron Spence}
\affiliation{Univ. Grenoble Alpes, CNRS, Grenoble INP, Institut N\'eel, 38402 Grenoble, France}

\author{Matthieu Dartiailh}
\affiliation{Univ. Grenoble Alpes, CNRS, Grenoble INP, Institut N\'eel, 38402 Grenoble, France}

\author{Baptiste Jadot}
\affiliation{Univ. Grenoble Alpes, CEA, Leti, F-38000 Grenoble, France}

\author{Emmanuel Chanrion}
\affiliation{Univ. Grenoble Alpes, CNRS, Grenoble INP, Institut N\'eel, 38402 Grenoble, France}

\author{Vivien Thiney}
\affiliation{Univ. Grenoble Alpes, CNRS, Grenoble INP, Institut N\'eel, 38402 Grenoble, France}

\author{Renan Lethiecq}
\affiliation{Univ. Grenoble Alpes, CNRS, Grenoble INP, Institut N\'eel, 38402 Grenoble, France}

\author{Benoit Bertrand}
\affiliation{Univ. Grenoble Alpes, CEA, Leti, F-38000 Grenoble, France}

\author{Heimanu Niebojewski}
\affiliation{Univ. Grenoble Alpes, CEA, Leti, F-38000 Grenoble, France}

\author{Christopher B{\"a}uerle}
\affiliation{Univ. Grenoble Alpes, CNRS, Grenoble INP, Institut N\'eel, 38402 Grenoble, France}

\author{Maud Vinet}
\affiliation{Univ. Grenoble Alpes, CEA, Leti, F-38000 Grenoble, France}

\author{Yann-Michel Niquet}
\affiliation{Univ. Grenoble Alpes, CEA, IRIG, 38000 Grenoble, France}

\author{Tristan Meunier}
\affiliation{Univ. Grenoble Alpes, CNRS, Grenoble INP, Institut N\'eel, 38402 Grenoble, France}

\author{Matias Urdampilleta	}
\email{matias.urdampilleta@neel.cnrs.fr}
\affiliation{Univ. Grenoble Alpes, CNRS, Grenoble INP, Institut N\'eel, 38402 Grenoble, France}

\date{\today}% It is always \today, today,
\begin{comment}
\begin{abstract}
Silicon-based spin qubits are among the leading candidates for a scalable quantum computer. To unveil their full potential regarding scalability and co-integration with classical electronics, Complementary-Metal-Oxide-Semiconductor (CMOS) processes are a promising route to achieve such challenges.
%We present the fabrication of qubit devices made with an industrial CMOS process followed by integration of micromagnets in the back end of line (BEOL) using processes at the single die level.
However, some flexibility in the back-end-of-line (BEOL) processes is needed to add new functionalities such as micromagnets or superconducting circuits to study the physics of these devices or achieve proof of concepts in addressing qubits - once the process stabilized it can be incorporated in the foundry-compatible process flow.
Here, we study a single electron spin qubit in a CMOS device with a micromagnet integrated in the flexible BEOL.
We exploit the synthetic spin orbit coupling (SOC) to control the qubit with electric field. Moreover, we investigate the spin-valley physics in the presence of SOC and we show an enhancement of the Rabi frequency at the spin-valley hotspot.
Finally, we probe the high frequency noise in the system using dynamical decoupling pulse sequences and demonstrate that charge noise dominates the qubit decoherence in this range.
\end{abstract}
\end{comment}

% Alternative Abstract Bernhard

%\begin{comment}
\begin{abstract}
For semiconductor spin qubits, complementary-metal-oxide-semiconductor (CMOS) technology is the ideal candidate for reliable and scalable fabrication.
Making the direct leap from academic fabrication to qubits fabricated fully by industrial CMOS standards is difficult without intermediate solutions.
With a flexible back-end-of-line (BEOL) new functionalities such as micromagnets or superconducting circuits can be added in a post-CMOS process to study the physics of these devices or achieve proof of concepts.
Once the process is established it can be incorporated in the foundry-compatible process flow.
Here, we study a single electron spin qubit in a CMOS device with a micromagnet integrated in the flexible BEOL.
We exploit the synthetic spin orbit coupling (SOC) to control the qubit via electric field and we investigate the spin-valley physics in the presence of SOC where we show an enhancement of the Rabi frequency at the spin-valley hotspot.
Finally, we probe the high frequency noise in the system using dynamical decoupling pulse sequences and demonstrate that charge noise dominates the qubit decoherence in this range.
\end{abstract}
%\end{comment}

\maketitle

\section{Introduction}

Foundry fabricated CMOS quantum dots offer a great opportunity to build high quality spin qubit devices with the prospect of creating a scalable quantum computer \cite{Gonzalez-Zalba2021, Veldhorst2017}. 
Many building blocks such as single-shot detection of electron spin states \cite{PRXQuantum.2.010353,PRXQuantum.3.040335}, electron manipulation in small arrays \cite{PhysRevApplied.14.024066,Gilbert2020, Ansaloni2020} or single qubit operation using electron spin resonance \cite{Zwerver2022} have been successfully implemented.
However, making the direct leap from academic fabrication to qubits fabricated fully by industrial CMOS standards is difficult without intermediate solutions.
An important step is to process all the parts of the qubits that are compatible with CMOS technology on an industrial level.
Only adding the components that are not yet compatible with large scale integration by post-CMOS processing in an academic clean room. 
This approach allows exploring the physics of these spin qubits and characterize new modules before their integration.
Among the different modules, the development of superconducting resonators integrated in the post-CMOS process has proven to be successful in the coupling of photon with spin qubits \cite{Yu2023}. % or for the readout of a quantum dot \cite{ElHomsy2023}.
Following the same principle, fabrication of magnetic materials for electric-dipole spin resonance (EDSR) \cite{Pioro-Ladriere2008,Kawakami2014, Leon2020, PhysRevApplied.11.061006, PhysRevApplied.15.044042, Yoneda2018} could also be integrated in the BEOL.
%, limiting the diffusion of magnetic material within the CMOS qubit layer.

In this paper, we study the physics related to the electrical control of a single-electron spin qubit using EDSR. The device is fabricated using a hybrid optical/e-beam lithography process to achieve the proof-of-concept of micromagnet integrability. We exploit the SOC to drive the qubit coherently and extract its coherence properties. Moreover, we show how the presence of valleys gives rise to a second driving mechanism thanks to spin-valley mixing combined with the synthetic SOC. Finally, we use dynamical decoupling to investigate the noise source at high frequency.
%\end{comment}
%%%%%%%%%%%%%%%%%%%%%%%%%%%%%%%%%%%%%%%%%%%%%%%%%%%%%%%%%%%%%%%%%%%%%%%%
\section{Device fabrication and operation}
\begin{figure}%
\includegraphics[width=\columnwidth]{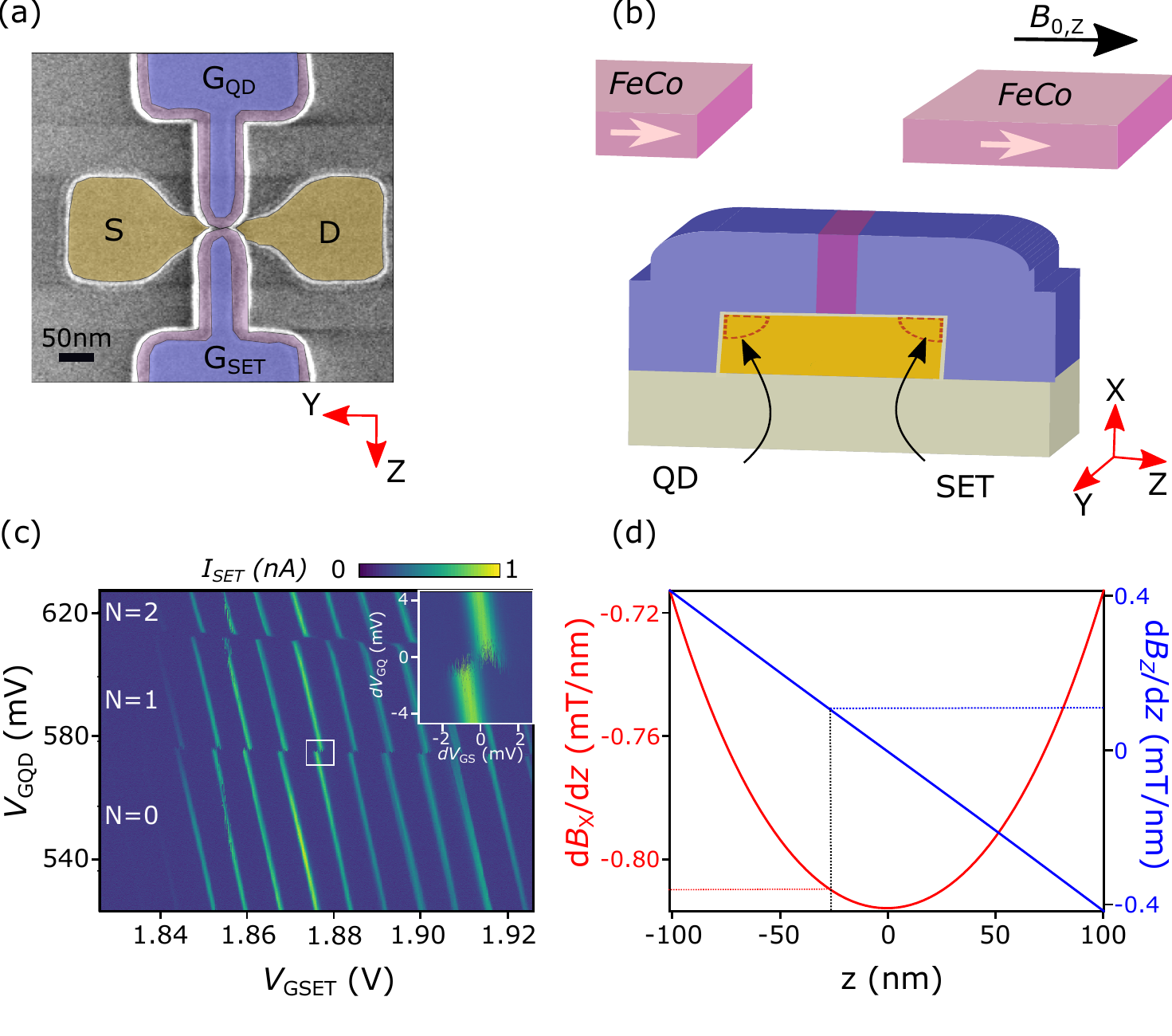}
\caption{
(a) False-coloured SEM image of a CMOS quantum dot device similar to the one used in this work. A pair of split gates (purple) is deposited on top of the channel (yellow) and SiN spacers (dark pink) are created to protect the channel from in-situ doping of the reservoirs.
(b) Cut through the device along the gates, showing the position of the SET and the QD. The micromagnet added on top is magnetized by the external magnetic field $B_0$.
(c) Stability diagram in the few electron regime for the quantum dot formed under gate $G_\text{QD}$. The current flowing through the quantum dot formed under gate $G_\text{SET}$ is used for charge sensing. An abrupt shift in the conductance peak corresponds to an electron entering or leaving the QD. In the inset a zoom on the transition used for the following measurements is shown. 
(d) Simulation of the magnetic gradients from the micromagnet. The transverse gradient $dB_x/dz$ (red) and the longitudinal gradient $dB_z/dz$ (blue) are plotted against the position across the nanowire. The gray line represents the estimated position of the QD.
}
\label{fig:device}%	
\end{figure}

The device, presented in Fig.\ref{fig:device}(a) is based on a \SI{80}{\nano\meter} wide silicon nanowire transistor from fully depleted silicon-on-insulator (FD-SOI) technology\cite{9830352}.
On top of a \SI{10}{\nano\meter} thick nanowire, \SI{6}{\nano\meter} SiO$_\textrm{2}$ gate oxide and \SI{5}{\nano\meter} TiN / \SI{50}{\nano\meter} polysilicon gate metal are deposited.
A single pair of split gates with a gate length and gate separation of \SI{50}{\nano\meter} is patterned using optical and electron-beam lithography (EBL).
SiN spacers are created to protect the device during the in situ doping of the reservoirs with phosphorus.
Before the device is post-processed in an academic clean room, it is buried in \SI{200}{\nano\meter} SiO$_\textrm{2}$ with tungsten vias to contact the gates and reservoirs.

In the post-CMOS process, the alignment with respect to the tungsten vias is made by depositing a set of markers and measuring the misalignment using scanning electron microscopy (SEM).
Contact is established with Ti/Al metallic lines using EBL and standard lift-off techniques.
The contact is ensured by an argon ion beam milling step prior to Ti/Al  deposition.
The Ti/Al lines are isolated from the \SI{300}{\nano\meter} thick FeCo micromagnet, depicted in Fig.\ref{fig:device}(b), by \SI{12.5}{\nano\meter} Al$_\textrm{2}$O$_\textrm{3}$ grown with atomic layer deposition (ALD).
Finally, the device is encapsulated with \SI{5}{\nano\meter} ALD-grown Al$_\textrm{2}$O$_\textrm{3}$ and a post-fabrication annealing is performed.

We use FeCo micromagnets with a $\SI{450}{\nano\meter}$ gap , inducing a magnetic gradient of the field component along an externally applied magnetic field ("longitudinal gradient"), and a magnetic gradient of the field components orthogonal to this external field ("transverse gradient").
The longitudinal gradient, shown in Fig.\ref{fig:device}(d), allows for individual frequency detuning \cite{Philips2022}. 
While useful for individual qubit addressability, it also opens the door to dephasing \cite{Culcer2009, Struck2020}.
In the current experiment, due to misalignment during the fabrication process observed by SEM, the spin qubit quantum dot is slightly shifted away from zero longitudinal gradient, see Fig.\ref{fig:device}(d).
It results in a finite dephasing gradient in the range of few $\SI{}{\mega\hertz /nm}$ when the electron is moved by charge noise along the quantization axis z.
Simulations of the micromagent (see Suppl. Mat. 2) yield a transverse gradient in the range of $\SI{0.8}{\milli T/nm}$ and it is limited by the encapsulation layer between the active region of the device and the micromagnet ($\SI{250}{\nano\meter}$).
The estimated stray field induced by the micromagnet is in the range of $\SI{160}{\milli\tesla}$, as will be confirmed by Larmor spectroscopy.

Each of the two gates enables the accumulation of electrons in a so-called corner dot.
The qubit dot (QD) contains a single electron whose spin is used as a qubit, and the other dot is operated as a single-electron transistor (SET) in the many electron regime \cite{PhysRevApplied.17.034047, PRXQuantum.2.010353}.
To load a single electron in the QD, we start by characterizing the stability diagram of the QD-SET system as a function of the gate voltages $G_\text{QD}$ and $G_\text{SET}$, see Fig. \ref{fig:device}(c).
We can identify the different charge occupation regimes of the QD and perform real time measurements of tunneling events by sitting on a detector Coulomb peak at the $N=0$ to $N=1$ charge transition sampling data at $\SI{20}{\kilo\hertz}$ (see inset of Fig.\ref{fig:device}(c)).

%%%%%%%%%%%%%%%%%%%%%%%%%%%%%%%%%%%%%%%%%%%%%%%%%%%%%%%%%%%%%%%%%%%%%%%%
\section{Spin relaxation characterization}

\begin{figure}%
\includegraphics[width=\columnwidth]{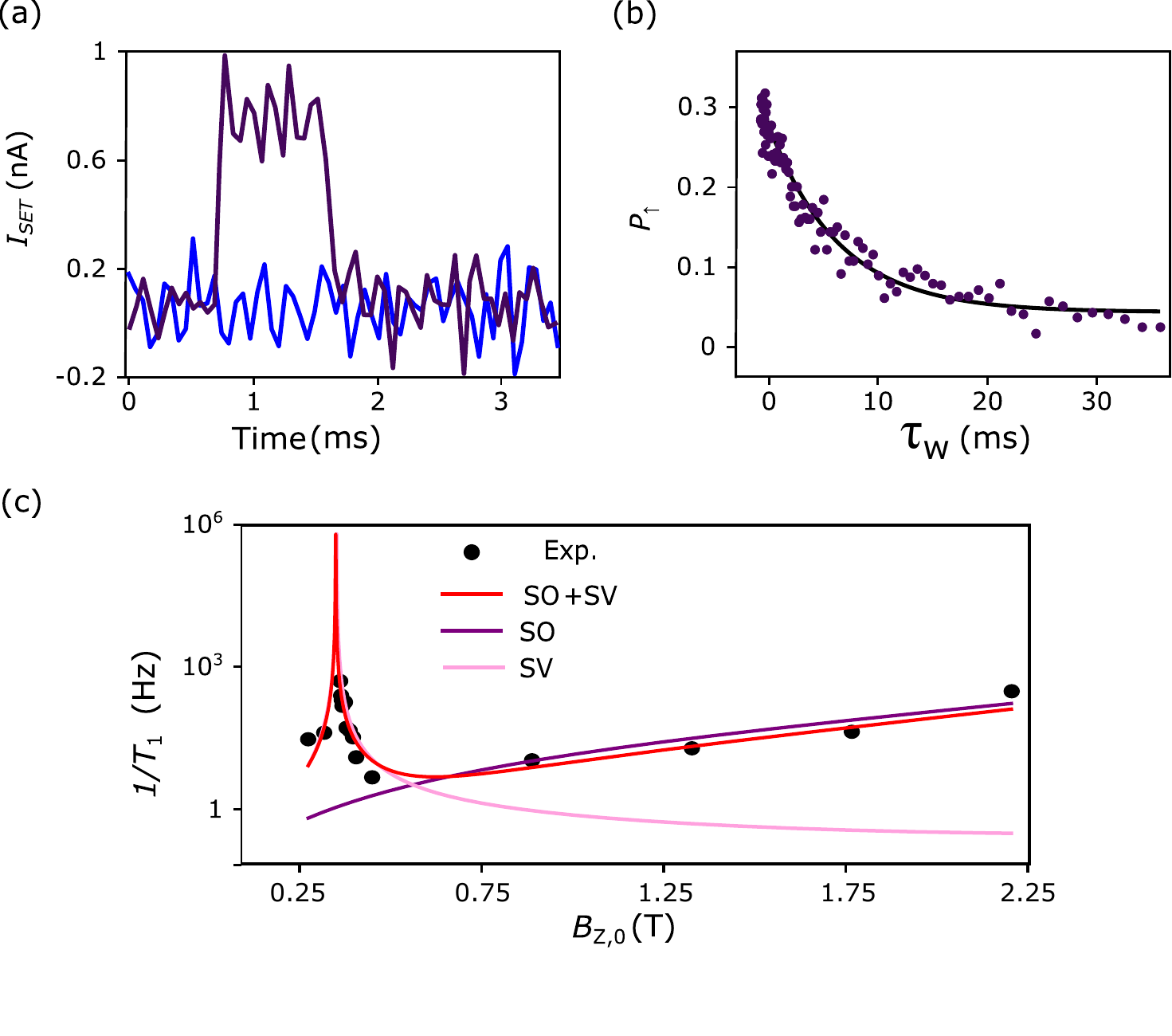}
\caption{
(a) Single shot traces representing typical scenarios of a spin-down electron and a spin-up electron being detected.
(b) Measurement of the spin relaxation time T$_\textrm{1}$ as a function of external applied magnetic field. An exponential decay is fitted to the data yielding a spin relaxation time of T$_\textrm{1}$ = $\SI{5.2}{ms}$. The initialization is performed by loading a random spin in the quantum dot.
(c) Spin relaxation rate 1/$\textrm{T}_\textrm{1}$ plotted for different total magnetic fields. At $\SI{0.36}{\tesla}$ the spin relaxation hotspot is visible. The data is fitted to determine the contribution of spin-orbit coupling (purple) spin-valley coupling (pink). The combination of spin-orbit and spin-valley coupling (red) fits the data best.
}
\label{fig:spin}%	
\end{figure}

Spin detection is performed by energy selective tunneling readout at finite magnetic field \cite{Elzerman2004, Morello2010}.
We start by emptying the qubit dot in the $N=0$ charge region, followed by a deep plunge in the $N=1$ charge state for a waiting time $\tau_\text{w}$.
We then move to the measurement position to the $N=0$ to $N=1$ transition.
We obtain a characteristic click observed in Fig.\ref{fig:spin}(a).
A current bump is observed when the loaded electron was in the spin up state and the current remains at the base level when a spin down electron was loaded.
To further characterize the readout configuration, we measure the probability to find the system in the spin up state deep in the $N=1$ charge state, as a function of $\tau_\text{w}$.
We obtain a characteristic relaxation curve, depicted in Fig.\ref{fig:spin}(b) from which we extract a relaxation time $T_1$ of $\SI{5.2}{\milli\second}$.
%The measurement gives a poor visibility of $75\%$, limited by the thermal broadening of the reservoir (estimated to $\SI{400}{\milli\kelvin}$) \cite{Keith2019}.
The measurement gives a limited visibility of $75\%$, and is further reduced in the following of the paper, as the magnetic field is decreased leading to a smaller $k_\text{B}T/E_z$ ratio, with $T$ the electron temperature (estimated to $\SI{400}{\milli\kelvin}$), $E_z=g\mu_\text{B}B_z$ the Zeeman energy, $k_\text{B}$ the Boltzmann constant, $g=2$ the g-factor of electrons in silicon, and $\mu_\text{B}$ the Bohr magneton \cite{Keith2019}. 

Probing relaxation allows to extract more information about the QD spectrum and its coupling to environment \cite{PhysRevLett.124.257701, Petit2018, Yang2013, PhysRevB.90.235315}.
In particular, we extract the valley splitting $E_\text{VS}=\SI{60}{\micro\electronvolt}$ by measuring $T_1$ as a function of magnetic field, see Fig.\ref{fig:spin}(c). It is worth noting that this value takes into account the micromagnet stray field (see Suppl. Mat. 2)
%Fitting the relaxation dependence with external field shows that the main mechanism is driven by Johnson-Nyquist noise for both contributions coming from the presence of excited orbitals ($SO^\text{JN}$, in purple) and valleys ($SV^\text{JN}$, in pink), see Suppl. Mat. 3 for the fitting parameters.
%The phonon contribution $SO^\text{Ph} + SV^\text{Ph}$, plotted in blue in Fig.\ref{fig:spin}(c), is negligible.
%This is in agreement with similar experiments where hotspot has been obtained in the same range and with identical dependence with magnetic field \cite{Borjans2019}.
%The strong contribution of the Johnson-Nyquist noise is attributed to the enhanced spin-orbit interaction due to the magnetic gradient combined with electric field fluctuations at the Larmor frequency \cite{Huang2021}.
%Also, the small spin-valley coupling extracted, $\Delta_\text{VS}=\SI{40}{\nano\electronvolt}$, can be explained by destructive interference between the intrinsic and extrinsic contributions to spin orbit coupling.

%%%%%%%%%%%%%%%%%%%%%%%%%%%%%%%%%%%%%%%%%%%%%%%%%%%%%%%%%%%%%%%%%%%%%%%%
\section{Electric-dipole spin resonance}

\begin{figure}%
\includegraphics[width=\columnwidth]{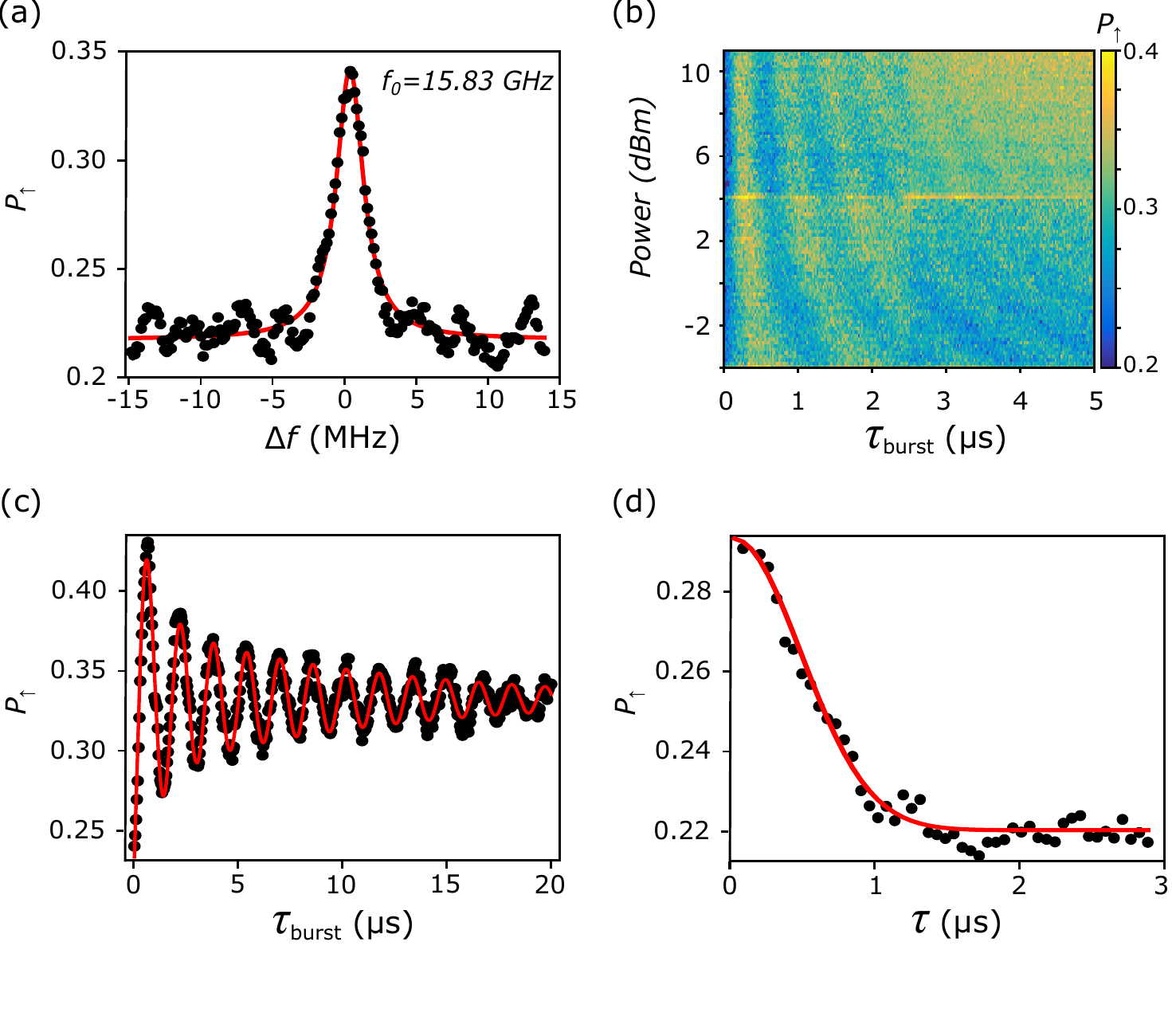}
\caption{
(a) EDSR spectrum of the single electron spin.
 Spin up state probability is measured as a function of the excitation frequency.
The peak is centered around the Larmor frequency $f_0$.
The initialization is performed by waiting $\SI{1}{\milli\second}$ at the measurement position where, in principle, only the ground state is accessible.
(b) Spin up probability as a function of the RF burst duration and the input power.
%(b) $f_0$ is measured as in (a) and plotted as a function of the external magnetic field.
(c) Rabi oscillations for $\SI{1}{\milli\volt}$ excitation amplitude.
The data points (black) are fitted using the $\cos(\omega t+\pi/4)/\sqrt{t}$ function to account for the slowly fluctuating nuclear field.
(d) Echo envelope measured using a Hahn echo sequence leading to an estimated T$_2^*$ of $\SI{500}{\nano\second}$.
}
\label{fig:EDSR}%	
\end{figure}

We probe the spectrum of the spin states by performing pulsed measurements using EDSR.
The EDSR drive is applied to $G_\text{QD}$ with a $\SI{1}{\milli\volt}$ amplitude.
%  We start with a coarse scan of excitation frequency using chirp pulses with $\SI{5}{\mega\hertz}$ frequency sweep and $\SI{10}{\micro\second}$ duration for $\SI{1}{\milli\volt}$ driving amplitude, after which we probe the spin probability using single shot readout.
 % The frequency sweep is then refined with monochromatic pulse around the region where finite excited spin probability has been measured, see Fig.\ref{fig:EDSR}(a).
At fixed magnetic field, we extract the Zeeman energy from a fit of the Larmor resonance (Fig.\ref{fig:EDSR}(a)). The Larmor frequency is then studied with respect to the external magnetic field, see Suppl. Mat. 2.
From such measurement, taking into account that $g=2$, the micromagnet stray field is evaluated to be in the range of $\SI{160}{\milli\tesla}$.
However, the magnetization of the micromagnet is changing with the applied external magnetic field leading to different Larmor frequencies depending on how the external magnetic field has been swept previously.
Tracking the Larmor frequency as a function of the external magnetic field allows us to reconstruct its micromagnet minor hysteresis loop presented in Suppl. Mat. 1 in agreement with a partially saturated magnet.

We now move to the coherent manipulation of the qubit.
We perform Rabi oscillations by sweeping the duration of a $\SI{1}{\milli\volt}$ AC drive at the Larmor frequency $f_0$.
We obtain the data presented in Fig.\ref{fig:EDSR}(b) and (c).
We observe a non-exponential decay of the Rabi oscillations which is attributed to the non-Markovian nature and slow fluctuation of the nuclear spin bath over the spin rotation time scale \cite{PhysRevLett.99.106803}.
The Rabi frequency over voltage drive ratio corresponds to approximately $\SI{1}{\mega\hertz}$ per $\SI{}{\milli\volt}$.
It is worth noting that the drive amplitude is limited to $\SI{1}{\milli\volt}$ in the present sample due to a heating effect, reducing drastically the visibility.

Using the above calibrated rotations we can perform a Hahn echo sequence to extract the intrinsic dephasing time $T_2^*$ and the Hahn-Echo coherence time $T_2^\text{E}$.
$T_2^*$ is directly extracted from the envelope of the echo, \cite{Pla2012}, as depicted in Fig.\ref{fig:EDSR}(d) and falls in the $\SI{500}{\nano\second}$ range when measured over few tens of minutes.
The small difference with reported values in natural silicon, \cite{Takeda2023, Kawakami2014} is explained by a smaller size of quantum dot leading to a shorter hyperfine-induced dephasing time.
The Hahn echo time $T_2^\text{E}$ is longer by two orders of magnitude, on the order of $\SI{36}{\micro\second}$, showing that the quasistatic noise due to hyperfine interaction has been refocused.

%%%%%%%%%%%%%%%%%%%%%%%%%%%%%%%%%%%%%%%%%%%%%%%%%%%%%%%%%%%%%%%%%%%%%%%%
\section{Valley enhanced EDSR}
\begin{figure}%
\includegraphics[width=\columnwidth]{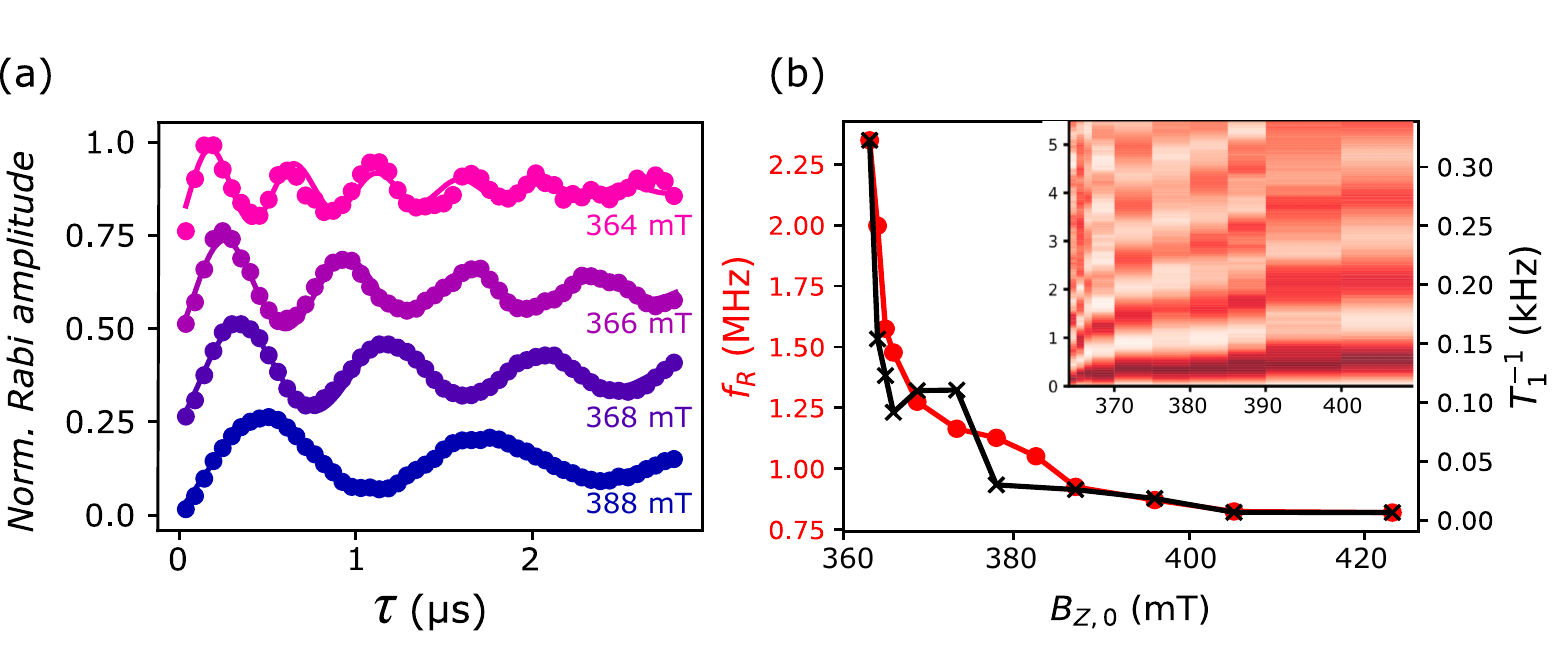}
\caption{
(a) Rabi oscillations at different magnetic fields for a constant drive amplitude ($V_{AC}=\SI{1}{\milli\volt})$.
(b) Rabi frequency and relaxation rate as a function of the external applied magnetic field. The relaxation rate indicates the presence of a hotspot close to $\SI{0.36}{\tesla}$. The Rabi frequency follows the same trend as the relaxation rate by increasing in the vicinity of this hotspot. Inset: colormap of Rabi oscillations as a function of field. 
}
\label{fig:Rabi_hotspot}%	
\end{figure}

To investigate the interplay between valley mixing and SOC we perform EDSR as a function of the magnetic field \cite{PhysRevB.95.075403, Corna2018}. 
%To rule out the contribution coming from variations in input power, the driving power is calibrated using two methods (see Suppl. Mat.): bare calibration of the RF lines and power induced excitation of the spin population.
Fig.\ref{fig:Rabi_hotspot}(a) presents Rabi oscillations performed at different magnetic fields. It shows an increase of the Rabi frequency as the field gets closer to the valley hotspot. 
This is supported by Fig.\ref{fig:Rabi_hotspot}(b) which plots both the Rabi frequency and the relaxation rate as a function of the magnetic field.
It is worth noting that measurement further on the hotspot is prevented by fast relaxation during readout.
This increase of EDSR Rabi frequency at the hotspot is related to the presence of a second drive mechanism which involves the presence of valley mixing in the silicon QD combined with synthetic SOC \cite{Huang2021}. 
More precisely, the microwave electric field allows a transition from two different valleys but same spin ($\ket{\textrm{v}_\textrm{-},\downarrow}$ and $\ket{\textrm{v}_\textrm{+},\downarrow}$) and the synthetic SOC couples the two opposite spins in different valleys ($\ket{\textrm{v}_\textrm{-},\uparrow}$ and $\ket{\textrm{v}_\textrm{+},\downarrow}$) which eventually leads to an opposite spins and same valley transition ($\ket{\textrm{v}_\textrm{-},\uparrow}$ to $\ket{\textrm{v}_\textrm{-},\downarrow}$) \cite{PhysRevB.97.155433}.

The Rabi frequency monotonously increases with the valley mixing which indicates that the two mechanisms are adding up resulting in a larger transverse field in the rotating frame.
A more quantitative estimation of the amplitude and phase of the two components of the driving fields in the rotating frame is impractical here.
It would require to change both the direction of the synthetic SOC field and the external magnetic field $B_z$ to map and disentangle the evolution of the two mechanisms in space.
While the presence of valleys offers an enhancement of EDSR driving speed it also increases the susceptibility of the qubit to charge noise.
We observe a similar trend between the Rabi frequency and the decoherence rate of Rabi oscillations, see Suppl. Mat. 7.
Though, we do not observe a clear sweetspot in quality factor as was proposed in ref. \cite{Huang2021}. 
We therefore need to understand further the noise spectrum of the qubit and in particular the influence of charge noise.

%%%%%%%%%%%%%%%%%%%%%%%%%%%%%%%%%%%%%%%%%%%%%%%%%%%%%%%%%%%%%%%%%%%%%%%%
\section{Probing high frequency charge noise using dynamical decoupling }

\begin{figure}%
\includegraphics[width=\columnwidth]{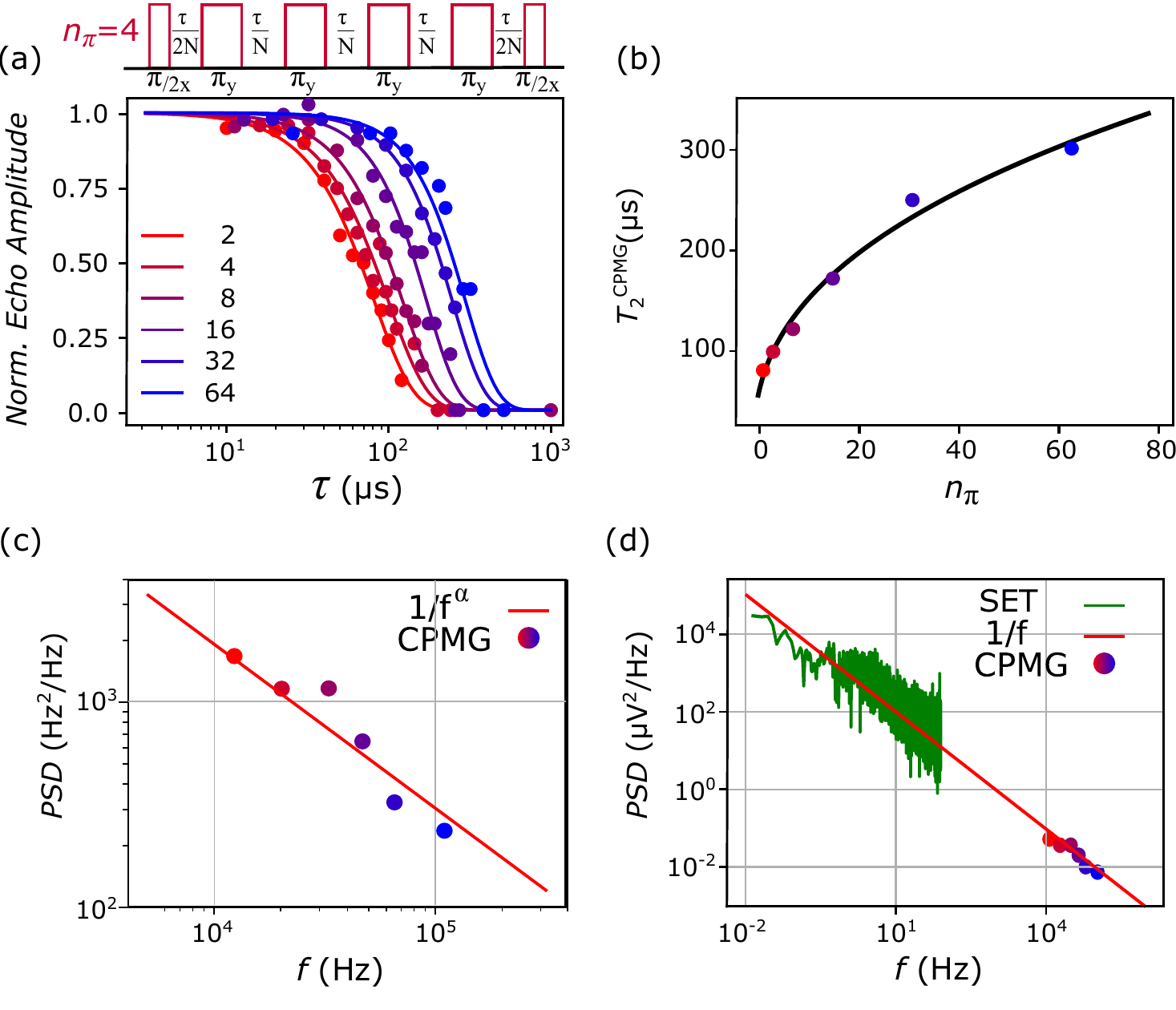}
\caption{
(a) Normalized echo amplitude as a function of total free evolution time using CPMG sequences. These sequences are composed of a series of N $\pi_y$ pulses, as depicted on the top of the figure for N=4 and finished with a $\pi_x/2$ probe at different timings $\tau+\delta\tau$, where $\delta\tau$ is swept to capture the whole echo. For each CPMG the echo is renormalized using the reference echo amplitude for $\tau$= $\SI{1}{\micro\second}$.
The different decay curves are fitted using the following expression $A(t)=exp{[-\left(t/T_2\right)^\beta]}$ with $\beta=1+\alpha$.
(b) Evolution of CPMG-enhanced T$_2$ with number of refocusing pulses. T$_2$ values are extracted from the decay curves in (a) and fitted using T$_2(n_\pi)\propto n_\pi^{\frac{\alpha}{1+\alpha}}$.
(b) Noise power spectral density of the qubit energy fluctuation for the evolution time corresponding to the CPMG $T_2^{n_\pi}$ times. Treating the CPMG sequence as a band pass filter allows to extract the noise PSD around the pulse repetition frequency. The $1/f$ trend, characteristic of charge noise, is plotted along the data points.
(c) Noise power spectral density in terms of effective gate voltage. The low frequency data points are extracted from time domain current measurements on the flank of a SET Coulomb peak followed by Fourier transform. The high frequency points correspond to data points from (a), turned into voltage fluctuations knowing the longitudinal magnetic gradient and the charge displacement due to gate voltage. 
}
\label{fig:CPMG}%	
\end{figure}

To probe noise at high frequency, we implement standard Carr-Purcell-Meiboom-Gill (CPMG) pulse sequences, which consist in applying multiple $\pi$ pulses instead of just one echo pulse in a given evolution time \cite{PhysRev.94.630, Meiboom1958, PhysRevA.58.2733}. Such a pulse sequence is characterized by the number of $\pi$ pulses $n_\pi$, the total free evolution time $\tau$, and the duration of a $\pi$ pulse $t_\pi$. The total time of one CPMG experiment is then given by $t = \tau + n_\pi t_\pi$. The amplitudes obtained after a final refocusing $\pi/2$ pulse are summarized in Fig.\ref{fig:CPMG}(a).
For $n_\pi$ $\pi$ pulses, the echo amplitude $A(t)$ is fitted using the following equation \cite{kawakami2016}:
  \begin{equation}
        A^{n_\pi}(t)=\exp{\left[-(t/T_2^{n_\pi})^{1+\alpha}\right]}
  \end{equation}
where $T_2^{n_\pi}$ is the characteristic CPMG time for information loss, for a given $n_\pi$. The parameter $\alpha$ is the noise color.

Another way to represent this dependence is to plot the CPMG-enhanced $T_2^{n_\pi}$ with respect to $n_\pi$, see Fig.\ref{fig:CPMG}(b).
The pulse number dependence follows a power law $n_\pi^{\alpha/(\alpha+1)}$ which in our case is consistent with a noise color of around $\alpha = 0.8$ over the frequency range of CPMG pulse repetitions ($10^4-10^6$Hz). This is slightly different from typical $\alpha=1$ in semiconductor structures \cite{Yoneda2018, Connors2022}, indicating that a finite ensemble of two-level systems in the surroundings of the qubit dominates the noise at higher frequencies. It is worth noting that the number $n_\pi$ of $\pi$ pulses is limited due to the poor gate fidelity induced by the large decoherence rate (see Suppl. Mat. 4).

A complementary analysis of CPMG data consists in considering these pulse sequences as tools for noise spectroscopy \cite{Yoneda2018, Connors2022}. We are interested in the phase information of the electron spin, at times $t$, which we can write as: $ \langle \exp\left[i\phi(t)\right] \rangle \equiv \exp\left[ -\chi(t) \right]$, where the phase decay envelope $\chi$ can be expressed as:

\begin{equation}
\label{chi}
\chi(t) = 4\pi^2 \int_0^{+\infty} df S'_{f_L}(f) \times W(f,\tau,t_\pi,n_\pi)
\end{equation}

Suppl. Mat. 6 gives details on this expression and on this section, following closely Connors et al. \cite{Connors2022}. In this expression, $S'_{f_L}(f)$ is the double-sided power spectral density (PSD) of the noise on the Larmor frequency, and $W(f,\tau,t_\pi,n_\pi)$ is the spectral weighting function corresponding to a specific pulse sequence. The normalized amplitude $A^{n_\pi}(\tau)$ plotted in Fig.\ref{fig:CPMG}(a) embodies the loss of phase information in the corresponding CPMG experiment, giving:
\begin{equation}
\label{link_chi_cpmg}
\chi(t = \tau + n_\pi t_\pi) = -\ln\left[ A^{n_\pi}(\tau,t_\pi,n_\pi) \right]
\end{equation}

If we assume a single-sided PSD of the form $S_{f_L}(f) = 2S'_{f_L}(f) = C/f^{\alpha}$, with $\alpha=0.8$ found in the last section, we can compute $C$ for all of our CPMG experiments using equation (\ref{chi}). We get $C^{n_\pi}$ values by averaging the results for $C$ at a given $n_\pi$, taking into account the normalized CPMG amplitudes between 0.15 and 0.85, where the amplitude is less likely to come from instrumental errors. In order to visualize these results, we note that the spectral weighting function $W$ peaks at $f=n_\pi D / 2\tau$, where $D=\tau / (\tau + n_\pi t_\pi)$ is the duty cycle of the pulse sequence. As a set of CPMG experiments at fixed $n_\pi$ is characterized by the time $T_2^{n_\pi}$, we chose to plot $ S_{f_L}(f^{n\pi} = n_\pi D / 2T_2^{n_\pi}) = C^{n_\pi}/(f^{n_\pi})^\alpha $ in Fig.\ref{fig:CPMG}(c).
 
To confirm the origin of charge noise as the main source of decoherence at high frequency, we now turn the PSD $S_{f_L}$ into a PSD $S_v$ in terms of fluctuations of gate voltage: we mimic the effect of this charge noise by an effective noise on the gate voltage. This voltage noise translates into qubit frequency fluctuations by moving the electron along the $z$ axis combined with a longitudinal gradient $\frac{dB_z}{dz}$.
It summarizes with the following expression:
  \begin{equation}
      S_f = \frac{g\mu_\text{B}}{h} \times \frac{dB_z}{dz} \times \delta_e \times S_v
      \label{eq:S}
  \end{equation}
where $h$ is the Planck constant and $\delta_e$ is the electron displacement along $z$ due to gate voltage on $G_\text{QD}$.
This displacement per voltage can be extracted from the Rabi oscillations assuming:
    \begin{equation}
      \delta_e= \frac{f_\text{R}}{V_\text{AC}\frac{g\mu_\text{B}}{h}\frac{dB_x}{dz}}
      \label{eq:delta}
  \end{equation}
with $V_\text{AC}$ the excitation voltage to drive Rabi oscillations at frequency $f_\text{R}$, $\frac{dB_x}{dz}$ the transverse gradient.
Using equation (\ref{eq:delta}) and a Rabi frequency of $f_\text{R} = \SI{1}{\mega\hertz}$ for $V_\text{AC}=\SI{1}{\milli\volt}$, we obtain an electron displacement of $\SI{45}{\pico\meter}/\SI{}{\milli\volt}$.
Using equation (\ref{eq:S}) we plot in Fig.\ref{fig:CPMG}(c) the PSD in terms of fluctuations in gate voltage.
To compare with a standard charge noise measurement, we also add on the same plot the PSD extracted from SET fluctuations obtained by simply transforming SET current measurements recorded while sitting on the flank of a Coulomb peak. We obtain a good agreement between the high and low frequency data points falling on the $1/f$ trend. This pink noise behavior over 6 decades confirms the origin of the decoherence mechanism at high frequency to be charge noise.
  
 To compare with other spin qubit platforms, we can convert the PSD from Fig. \ref{fig:CPMG}(d) at $\SI{1}{\hertz}$ ($\SI{1000}{\micro\volt}^2/\SI{}{\hertz}$) to a PSD in terms of chemical potential using the lever arm ($\SI{0.25}{\electronvolt}/\SI{}{\volt}$). We obtain $\SI{8}{\micro\electronvolt}/\SI{}{\hertz^{1/2}}$. 
 This value is one to two orders of magnitude higher compared to commonly reported values in literature\cite{PhysRevApplied.14.024066,Connors2022, Kranz2020}, but similar to foundry fabricated spin qubits \cite{Zwerver2022}. In our case, this large charge noise amplitude points toward a degradation of device quality during the post-CMOS process which uses lower standard processes.
However, the high frequency fluctuations of the qubit are similar to what is reported in other platforms with less nominal charge noise while the longitudinal gradient is comparable. Therefore, we attribute the lower influence of charge noise to a small charge displacement susceptibility ($\frac{dr}{dV}= \SI{45}{\pico\meter}/\SI{}{\milli\volt})$ due to confinement in the corner of the nanowire.

%%%%%%%%%%%%%%%%%%%%%%%%%%%%%%%%%%%%%%%%%%%%%%%%%%%%%%%%%%%%%%%%%%%%%%%%

\begin{comment}
    \begin{figure}%
\includegraphics[width=\columnwidth]{FIG5.pdf}
% \includegraphics[width=\columnwidth]{fig1.pdf}%
\caption{
(a) Noise power spectral density of the qubit energy fluctuation for the evolution time corresponding to the CPMG $T_2^{n_\pi}$ times. Treating the CPMG sequence as a band pass filter allows to extract the noise PSD around the pulse repetition frequency. The $1/f$ trend, characteristic of charge noise, is plotted along the data points.
(b) Noise power spectral density in terms of effective gate voltage. The low frequency data points are extracted from time domain current measurements on the flank of a SET Coulomb peak followed by Fourier transform. The high frequency points correspond to data points from (a), turned into voltage fluctuations knowing the longitudinal magnetic gradient and the charge displacement due to gate voltage. 
}
\label{fig:PSD}%	
\end{figure}
\end{comment}

%%%%%%%%%%%%%%%%%%%%%%%%%%%%%%%%%%%%%%%%%%%%%%%%%%%%%%%%%%%%%%%%%%%%%%%%
\section{Conclusion}
In conclusion, we have seen that a post-CMOS process can be used to implement new functionalities to foundry-fabricated devices. With this process, we patterned a micromagnet on top of a CMOS device, which enabled coherent manipulation of an electron spin through EDSR. We identify two EDSR mechanisms: the one based on bare synthetic SOC and the one induced by combining the spin-valley mixing near the valley hotspot with the synthetic SOC.
We further investigate the coherent properties of the spin qubit which shows a decoherence time around $\SI{500}{\nano\second}$ due to hyperfine coupling to surrounding nuclear spins. Refocusing this low frequency noise through dynamical decoupling pulse sequences allowed to enhance the qubit coherence time by three orders of magnitude, and to provide evidence that high frequency noise has electrical origin, with an amplitude comparable to state of the art spin qubit realizations.
While the charge noise amplitude is relatively high, the induced decoherence at large frequency is relatively small, due to the strong confinement in the corner of the nanowire, at the cost of lower Rabi frequency.
Lowering charge noise by two orders of magnitude is at reach and would contribute to CMOS-based spin qubits promising prospects.

\section{Acknowledgment}
We acknowledge technical support from L. Hutin, D. Lepoittevin, I. Pheng, T. Crozes, L. Del Rey, D. Dufeu, J. Jarreau, C. Hoarau and C. Guttin. 
We thank S. De Franceschi and R. Maurand for fruitful discussions and I. De Moraes and N. Dempsey for help with micromagnet fabrication.
B.K., D.J.N. acknowledges the GreQuE doctoral programs (grant agreement No.754303). The device fabrication is funded through the Mosquito project (Grant agreement No.688539).
This work is supported by the Agence Nationale de la Recherche through the CRYMCO and the PEPR PRESQUILE project. This project receives as well funding from the project QuCube (Grant agreement No.810504) and the project QLSI (Grant agreement No.951852).

\section{Authors contributions}
H.N, B.B., and M.V. were responsible for the front-end fabrication of the device. B.K.  postprocessed the micromagnet and contacts on the CMOS device. B.K. performed measurement with help from V.E. and M.U.. V.E. performed the micromagnetic simulations and CPMG analysis. M.U., V.E. and B.K. co-wrote the manuscript with inputs from all the authors. T.M and M.U supervised and initiated the project.

\appendix

\section{Appendix 1 - Micromagnet minor hysteresis loop}
\begin{figure}%
\includegraphics[width=\columnwidth]{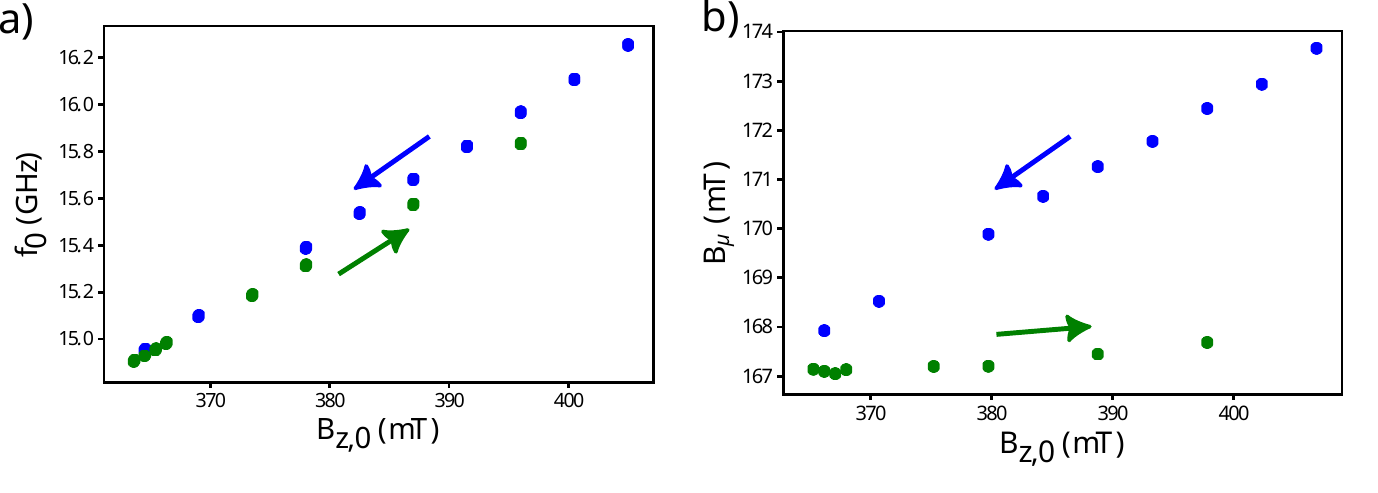}
\caption{a) Larmor frequencies plotted for different magnetic fields. b) Micromagnet magnetization depending on the external magnetic field. In both panels, the magnetic field is lowered monotonously for the blue points and it is increased monotonously for the green points.}
\label{fig:Hysteresis}%	
\end{figure}

%Hysteresis magnetic field
By taking measurements of the Larmor frequency at different magnetic fields, the hysteresis of the micromagnet is measured and its magnetization is estimated. Fig.\ref{fig:Hysteresis} (a) shows measurements of the Larmor frequency for the magnetic field being ramped up (green) and ramped down (blue). Taking into account a g-factor of 2, the magnetization of the micromagnet is estimated using eqn. (\ref{eq:Hysteresis}). The magnetization B$_{\mu}$ is the difference between the external magnetic field B$_\text{0,z}$ and the expected field from the Larmor frequency f$_\text{0}$.  
\begin{equation}
B_{\mu} = \frac{h f_0}{g \mu_B} - B_\text{0,z}
\label{eq:Hysteresis}
\end{equation}
where h is Planck's constant and $\mu_\text{B}$ is the Bohr magneton.
 %For an increasing external magnetic field the magnetization needs to remain constant or has to increase. The configuration where the magnetisation remains constant is plotted in Fig.\ref{fig:Hysteresis} (b). 
 The field generated by the micromagent is estimated to be around $\SI{160}{\milli\tesla}$. This result represents a lower bound for the magnetization at saturation of the micromagnet. 
 %As a hard magnet behavior is expected for FeCo at cryogenic temperatures \cite{tbf}, the choice of a constant magnetisation and a large hysteresis loop is justified.

\section{Appendix 2 - Micromagnetic simulations}

In order to simulate the magnetic field created by the micromagnet, we follow the development of ref.\cite{Yang1990}. We consider perfect parallelepiped-shaped magnets, as the impact of edge fluctuations on the magnetic field felt by the qubit is negligible. We also consider that the magnets have uniform magnetization along the $z$-axis.

The magnetic field produced by a single magnetic dipole reads:
\begin{equation}
    B_\text{dip}(r, \theta) = \frac{\mu_0}{4\pi} \times \frac{m}{r^3} \times \left( 2\cos(\theta)\overrightarrow{r} + \sin(\theta)\overrightarrow{\theta} \right) 
\end{equation}
with $\overrightarrow{r}$ and $\overrightarrow{\theta}$ the usual spherical unit vectors. Integrating this formula over space yields the following expression for the magnetic field created by a slab of magnetic material:

\begin{align}
    B_z(x,y,z) = -\frac{\mu_0m}{4\pi} (& F_1(x,y,-z) + F_1(-x,y,-z) \nonumber\\
    +& F_1(x,-y,-z) + F_1(-x,-y,-z) \nonumber\\
    +& F_1(x,y,z) + F_1(-x,y,z) \nonumber\\
    +& F_1(x,-y,z) + F_1(-x,-y,z) ) \nonumber\\
\end{align}

\begin{align}
    B_x&(x,y,z) = \frac{\mu_0m}{4\pi}\times \nonumber\\
    &\ln\left(\frac{F_2(y,-x,-z,D,W,L)F_2(y,x,z,D,W,L)}{F_2(y,x,-z,D,W,L)F_2(y,-x,z,D,W,L)}\right) \\
    B_y&(x,y,z) = \frac{\mu_0m}{4\pi}\times \nonumber\\
    &\ln\left(\frac{F_2(x,-y,-z,W,D,L)F_2(x,y,z,W,D,L)}{F_2(x,y,-z,W,D,L)F_2(x,-y,z,W,D,L)}\right)
\end{align}
with $(x,y,z)$ the Cartesian coordinates of some point outside the magnetic slab expressed in the frame with origin at the center of the slab, $m$ the saturation magnetization, $(2W,2D,2L)$ the dimensions of the slab along $(x,y,z)$, and $F_{1,2}$ the functions defined as follows:

\begin{align}
    F_1&(x,y,z) = \nonumber\\
    &\frac{(x+W)(y+D)}{(z+L)\sqrt{(x+W)^2+(y+D)^2+(z+L)^2}}\\
    F_2&(x,y,z,x',y',z') = \nonumber\\ 
    &\frac{\sqrt{(z+z')^2+(y+y')^2+(x'-x)^2} + x'-x}{\sqrt{(z+z')^2+(y+y')^2+(x'-x)^2} -x'-x}
\end{align}

The saturation magnetization of FeCo highly depends on the slab history and stoichiometry. We adjust it thanks to equation (\ref{eq:Hysteresis}) from previous section, based on Larmor spectroscopy.

\section{Appendix 3 - Spin relaxation}
%Fitting Hotspot, describe the procedure to make the different fit as well as the equations and numerical values

The spin-valley relaxation hotspot, as well as the overall magnetic field dependence of the spin relaxation rate is modeled by adding up the contributions of spin-valley (SV) and spin-orbit (SO) couplings for both Johnson–Nyquist (JN) noise and phonons (PH) following Ref. It is worth noting that in the following we only consider contribution from the intrinsic SOC \cite{PhysRevB.90.235315}. 
\begin{equation}
\frac{1}{T_1} = \Gamma_{SV,JN} + \Gamma_{SV,Ph} + \Gamma_{SO,JN} + \Gamma_{SO,Ph}
\label{eqn:rates}
\end{equation}
The relaxation rate induced by spin-valley coupling is described by eqns. (\ref{eqn:Gamma_{SV,JN}}) and (\ref{eqn:Gamma_{SV,Ph}}) and corresponds to the pink curve in Fig.\ref{fig:spin}(c).
\begin{align}
\label{eqn:Gamma_{SV,JN}}
\Gamma_{SV,JN} &= C_{SV,JN}^2 \left(\frac{E_Z}{E_0}\right) \frac{\exp(\frac{E_Z}{k_bT})+1}{\exp(\frac{E_Z}{k_bT})-1} F\left(E_{VS}, \Delta \right) \\
\label{eqn:Gamma_{SV,Ph}}
\Gamma_{SV,Ph} &= C_{SV,Ph}^2 \left(\frac{E_Z}{E_0}\right)^5 \frac{\exp(\frac{E_Z}{k_bT})+1}{\exp(\frac{E_Z}{k_bT})-1} F\left(E_{VS}, \Delta \right) \\
\label{eq:F_large}
F\left(E_{VS}, \Delta_{VS} \right) &= \frac{2\Delta_{SV}^2(E_2-E_1)^2}{(E_1^2+\Delta_{SV}^2)(E^2_2+\Delta_{SV}^2)} \qquad \text{if } E_{VS} 	\geq E_Z \\
\label{eq:F_small}
F\left(E_{VS}, \Delta_{VS} \right) &= \frac{2(E_1E_2 + \Delta_{SV}^2) ^2}{(E_1^2+\Delta_{SV}^2)(E^2_2+\Delta_{SV}^2)} \qquad \text{if } E_{VS} < E_Z \\
\label{eq:E1}
E_1 &= E_{VS} - E_Z + \sqrt{(E_{VS} - E_Z)^2 + \Delta_{SV}^2} \\
\label{eq:E2}
E_2 &= E_{VS} + E_Z + \sqrt{(E_{VS} + E_Z)^2 + \Delta_{SV}^2}
\end{align}

with $C_{i,j}$ ( $i\in \{SV,SO\}$, $j\in \{JN,Ph\}$) being the respective coupling strengths, $E_\textrm{Z}$ the Zeeman energy, $E_\textrm{0}$ an arbitrary energy reference, $k_\textrm{b}$ the Boltzmann constant and $T$ the temperature. F is a function of the valley splitting E$_\textrm{VS}$ and the spin-valley half-gap $\Delta_\textrm{VS}$. It is responsible for modeling the spin-valley hot spot.
The valley hotspot occurs due to the anticrossing of the $\ket{\textrm{v}_\textrm{-},\uparrow}$, $\ket{\textrm{v}_\textrm{+},\downarrow}$ states. The valley splitting E$_\textrm{VS}$ determines the B field at which it appears.\\
As $E_\textrm{Z}$ is proportional to the magnetic field, the global dependence of each rate is given by the exponent of $\left(\frac{E_\textrm{Z}}{E_\textrm{0}}\right)$. 
%The Johnson–Nyquist noise contribution is linear while the phonon noise contribution scales with the power of five with $B$. However, far from the hotspot the contribution of spin-valley coupling is negligible as the spin valley effects strongly decrease.

The spin-orbit coupling is described by eqns. (\ref{eqn:Gamma_{SO,JN}}) and (\ref{eqn:Gamma_{SO,Ph}}) and corresponds to the purple curve. The Johnson–Nyquist noise contribution scales with the power of three while the phonon noise contribution scales with the power of seven with B.
\begin{align}
\label{eqn:Gamma_{SO,JN}}
\Gamma_{SO,JN} &= C_{SO,JN}^2 \left(\frac{E_Z}{E_0}\right)^3 \frac{\exp(\frac{E_Z}{k_bT})+1}{\exp(\frac{E_Z}{k_bT})-1} \\
\label{eqn:Gamma_{SO,Ph}}
\Gamma_{SO,Ph} &= C_{SO,Ph}^2 \left(\frac{E_Z}{E_0}\right)^7 \frac{\exp(\frac{E_Z}{k_bT})+1}{\exp(\frac{E_Z}{k_bT})-1}
\end{align}
The fits show that Johnson–Nyquist noise dominates over phonons in almost the entire range of B field, as can be seen in the red curve in Fig.\ref{fig:spin}(c). %In agreement with Ref. \cite{PhysRevApplied.17.034047} phonons start to become important at \SI{2.5}{\tesla}, as can be seen in the blue curve, where only the phonon contribution is plotted.

\section{Appendix 4 - $\pi$ pulse fidelity}
The fidelity of the $\pi$ pulses for the CPMG sequence is estimated by probing the evolution of the echo amplitude with the number of pulses. The time delay between each pulses is set to $\SI{2}{\micro\second}$ to allow the refocusing of quasistatic noise see Fig. \ref{fig:CPMG}(a).

\begin{figure}%
\includegraphics[width=75mm]{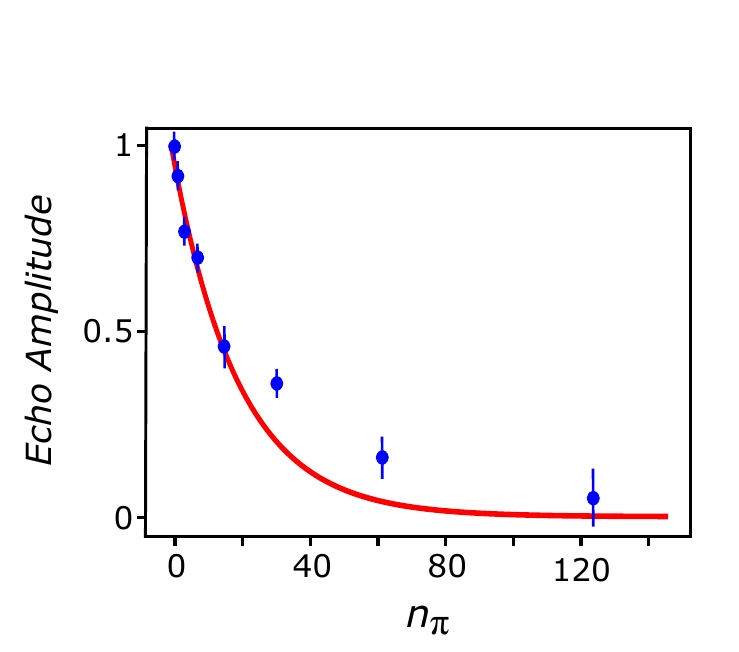}
\caption{
Echo amplitude as a function of the number of pulses $n_\pi$. The red curve is an exponential fit to the data.  
}
\label{fig:fid}%	
\end{figure}

\section{Appendix 5 - Hahn-Echo}

The simplest dynamical decoupling pulse sequence implemented here to refocus low frequency noise is the Hahn-Echo sequence. It consists in 1) Preparing the electron spin in its down state 2) Applying a $\pi/2$ pulse along the $X$ axis of the Bloch sphere to have a balanced superposition of up and down state with known phase 3) Waiting a variable time $\tau_\text{wait}$: the qubit slowly loses phase information 4) Applying a $\pi$ pulse along $X$ to refocus phase noise that accumulated slower that $\tau_\text{wait}$ 5) Waiting $\tau_\text{wait}+\Delta\tau_\text{wait}$ 6) Applying a $\pi/2$ pulse along $X$ again 7) Measure the spin state.

The $\Delta\tau_\text{wait}$ added in the last waiting time enables to measure an echo feature \cite{Pla2012}. If we apply the last $\pi/2$ pulse after exactly $\tau_\text{wait}$, we would measure the refocused signal, minus what has been lost due to higher frequency noise. But measuring this echo amplitude for variable $\Delta\tau_\text{wait}$ shows a broadening of the echo signal due to dephasing. We can then access $T_2^*$ by fitting the data to the following expression :
\begin{equation}
    P_\uparrow(\Delta\tau_\text{wait}) = B\times\exp\left[ - \frac{1}{2}\left(\frac{\Delta\tau_\text{wait}}{C}\right)^2 \right] + D
\end{equation}

\begin{figure}%
\includegraphics[width=\columnwidth]{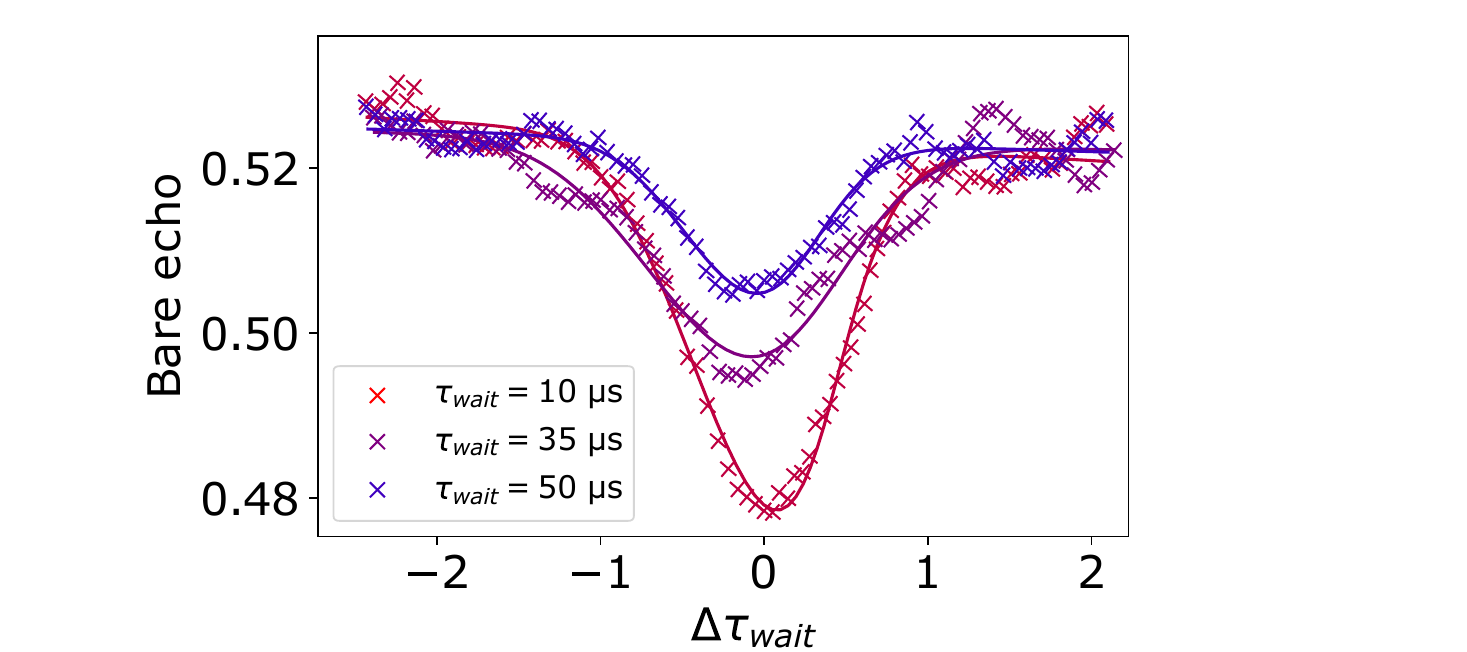}
\caption{Hahn-Echo final amplitude plotted for different waiting times $\tau_\text{wait}$
}
\label{fig:echo_traces}%	
\end{figure}

$T_2^*$ is then given by the full width at half maximum of this Gaussian-shaped curve: $T_2^* = 2\sqrt{2\ln{2}}\times C$. In Fig. \ref{fig:echo_traces}(a), we plotted this echo for $\tau_\text{wait} =  10, 35, \SI{50}{\micro\second}$. We note that between $\SI{3}{\micro\second}$ and $\SI{10}{\micro\second}$, the echo shows no visible degradation. This is due to the SET signal level that varies between experiment. This is why, for echo or CPMG experiments, before each shot of a specific pulse sequence, we first measure a shot of a reference Hahn-Echo with $\tau_\text{wait}=\SI{3}{\micro\second}$. The visibility of this reference echo is used to renormalize the subsequent shot visibility. In Fig.\ref{fig:echo_traces}(b), we plot the renormalized echo amplitude at $\Delta\tau_\text{wait}=0$, for different waiting times $\tau_\text{wait}$. The characteristic decay time of this curve is the Hahn-Echo time, $T_2^\text{E} = \SI{36}{\micro\second}$

\section{Appendix 6 - Analysis of CPMG data}

\begin{figure*}%
\includegraphics[width=\textwidth]{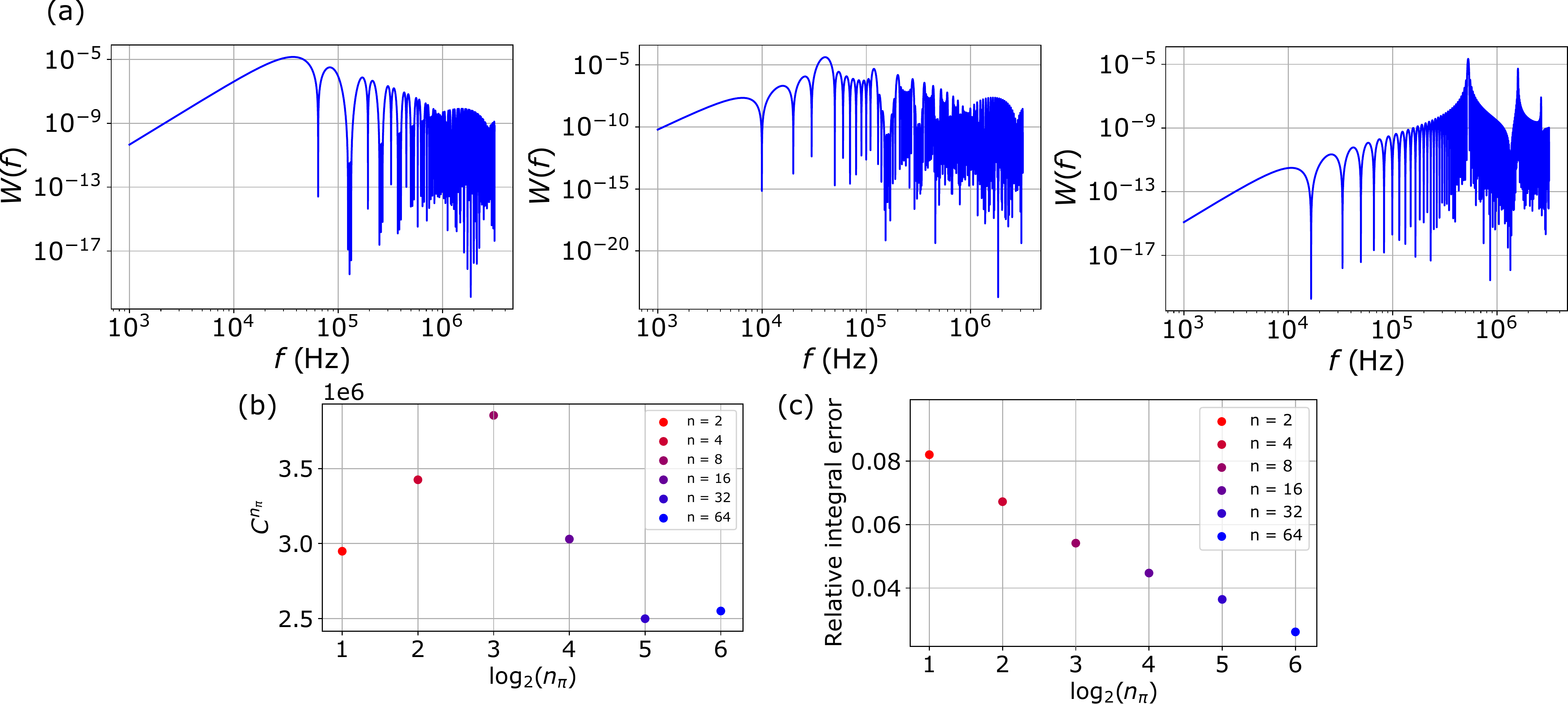}
\caption{(a) Spectral weight functions corresponding to CPMG sequences with $t_\pi=\SI{354}{\nano\second}$, and from left to right : $\tau = 30, 96, \SI{25.6}{\micro\second}$ and $n_\pi = 2, 8, 64$. (b) $C^{n_\pi}$ constants corresponding to assumed power spectral density model $S_{f_\text{L}}(f) = C/f^\alpha$. Each point is an average of these $C$'s computed from equation (\ref{Cs}) at fixed $n_\pi$. (c) Estimation of the error in numerical integration of relation (\ref{Cs}), computed by using the same numerical integration parameters for the integral of relation (\ref{error}) and comparing with expected result $\tau/2$.
}
\label{fig:Appendix_CPMG}%	
\end{figure*}

In this section, we develop on how to get power spectral density (PSD) data from CPMG experiments (main Fig.\ref{fig:CPMG}), following closely \cite{Connors2022}. The phase of the qubit during an experiment containing a pulse sequence is given by:

\begin{equation}
\phi(t) = 2\pi \int_{-\infty}^{+\infty} dt' f_\text{L}(t') \times y(t',\tau,t_\pi,n_\pi)
\end{equation} 
where $f_\text{L}$ is the Larmor frequency, and $y(t',\tau,t_\pi,n_\pi)$ describes the pulse sequence, characterized by the total free evolution time $\tau$, the $\pi$ pulse time $t_\pi$ and the number of pulses $n_\pi$.

We are interested in the time expectation value of the phase part of the qubit state. If we assume that fluctuations in the Larmor frequency $f_\text{L}$ are Gaussian, fluctuations in the qubit phase $\phi(t)$ are Gaussian too, and we can write:
\begin{equation}
\langle \exp\left[i\phi(t)\right] \rangle \equiv \exp\left[ -\chi(t) \right] = \exp\left[ \frac{-\langle\phi(t)^2\rangle}{2} \right]
\end{equation}
The Wiener-Khinchin theorem states that the double-sided PSD is the Fourier transform of the auto-correlation. Using this, and defining the spectral weighting function from the Fourier transform of $y$:
\begin{equation}
W(f,\tau,t_\pi,n_\pi) = |\hat{y}(f,\tau,t_\pi,n_\pi)|^2,
\end{equation}
we finally get for the decay envelope:
\begin{equation}
\label{chi_supp}
\chi(t) = 4\pi^2 \int_0^{+\infty} df S'_{f_\text{L}}(f) \times W(f,\tau,t_\pi,n_\pi)
\end{equation}

The analytic expression of $W$ is given in ref.\cite{Biercuk2009}. Suppl. Fig.\ref{fig:Appendix_CPMG}(a) shows the shape of $W$ for different pulse sequence parameters. We can see that for some parameters, $W(f,\tau,t_\pi,n_\pi)$ is strongly peaked around $f = n_\pi D / 2\tau$ with $D = \tau / (\tau + n_\pi t_\pi)$ is the duty cycle of the pulse sequence. This explains why we can use CPMG experiments to probe the qubit noise at specific frequencies. We also see that harmonics in $W$ can become relevant, so that we can't sum up the whole noise measured in a specific CPMG experiment to the peak frequency. In order to analyze in the same way all of our CPMG data, we proceed as follows. We assume the noise in the system is described by a single-sided PSD of the form $S_{f_\text{L}}(f) = C/f^\alpha$, where $\alpha$ has been fitted to 0.8 in the main text. Plugging this PSD in equation (\ref{chi_supp}), and remembering that $\chi(t)$ is linked to measured CPMG amplitudes through $\chi(t = \tau + n_\pi t_\pi) = -\ln\left[ A(\tau,t_\pi,n_\pi) \right]$, we get:
\begin{align}
\label{Cs}
C =& -\ln\left( A(\tau,t_\pi,n_\pi) \right) \times \nonumber \\
~& \left[ 4\pi^2  \int_0^{+\infty} df \frac{W(f,\tau,t_\pi,n_\pi)}{f^\alpha} \right]^{-1}
\end{align}
where we can compute the integral numerically. For this, we choose the upper bound of integration so that it encompasses the first three relevant harmonics of the spectral weight, given by odd multiples of the peak frequency. For each $n_\pi$ we average the resulting $C$'s to get $C^{n\pi}$ values used in main Fig.\ref{fig:CPMG}. The $C^{n_\pi}$'s are plotted in Suppl. Fig.\ref{fig:Appendix_CPMG}(b) and notably fall in the same order of magnitude region. This suggests that the simple $1/f^\alpha$ noise model is consistent on the relevant frequency range of CPMG experiments.

The confidence in numerical integration is estimated thanks to the following relation \cite{Connors2022}:
\begin{equation}
\label{error}
\int_0^{+\infty} df W(f,\tau,t_\pi,n_\pi) = \frac{\tau}{2}
\end{equation}
For each computation of $C$, we computed this integral of $W$ using the same integration parameters, with residual error coming mainly from the finite upper bound for integration. We plot the relative error in Suppl. Fig.\ref{fig:Appendix_CPMG}(c).

%TODO : check "higher n_pi, sharper peaks" for fixed rest
%TODO : check "higher n_pi, lower relative weight of harmonics", for fixed rest
\section{Appendix 7 - Rabi time near the hotspot}
\begin{figure}%
\includegraphics[width=\columnwidth]{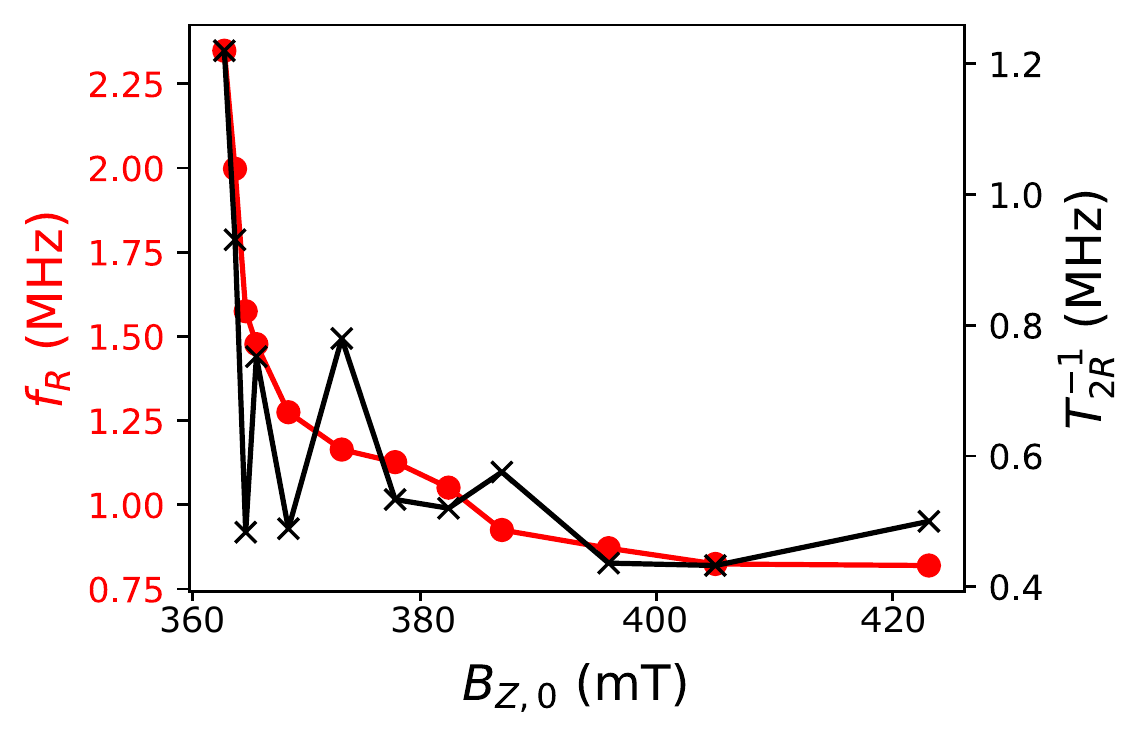}
\caption{Characteristic decay time of Rabi oscillations (black crosses) and Rabi frequency (red dots) near the hotspot.}
\label{fig:Rabi_T2}%	
\end{figure}

Fig.\ref{fig:Rabi_T2} presents the evolution of the damping of Rabi oscillations as a function of the magnetic field by fitting the Rabi oscillations presented in Fig.\ref{fig:EDSR} (a). 

%\section{Appendix 8 - Excitation power calibration}

%%%%%%%%%%%%%%%%%%%%%%%%%%%%%
\newpage 
\clearpage

\end{document}